# Comparison of Heparin Red, Azure A and Toluidine Blue assays for direct quantification of heparins in human plasma

U. Warttinger[1], C. Giese[1], Roland Krämer[1]

Correspondence to:
Roland Krämer, phone 0049 6221 548438, fax 0049 6221 548599
E-mail: kraemer@aci.uni-heidelberg.de

[1] Heidelberg University, Inorganic Chemistry Institute, Im Neuenheimer Feld 270, 60129 Heidelberg, Germany

## Abstract

Heparins are are sulfated polysaccharides that have tremendous clinical importance as anticoagulant drugs. Monitoring of heparin blood levels can improve patient safety. In clinical practice, heparins are monitored indirectly by their inhibtory effect on coagulation proteases. Drawbacks of these established methods have stimulated the development of simple direct detection methods with cationic dyes that change absorbance or fluorescence upon binding of polyanionic heparin. Very few such dyes or assay kits, however, are commercially and widely available to a broad community of researchers and clinicians. This study compares the performance of three commercial dyes for the direct quantification of unfractionated heparin and the widely used low-molecular-weight heparin enoxaparin. Two traditional metachromatic dyes, Azure A and Toluidine Blue, and the more recently developed fluorescent dye Heparin Red were applied in a mix-and-read microplate assay to the same heparin-spiked human plasma samples. In the clinically most relevant concentration range below 1 IU (international units) per mL, only Heparin Red is a useful tool for the determination of both heparins. Heparin Red is at least 9 times more sensitive than the metachromatic dyes which can not reliably quantify the heparins in this concentration range. Unfractionated heparin levels between 2 and 10 IU per mL can be determined by all dyes, Heparin Red being the most sensitive.

## Keywords

## Introduction

Heparins are widely used clinical anticoagulants, about one billion doses are applied each year. Heparin is a linear, polydisperse polysaccharide consisting of disaccharide repeating units (scheme 1) and has a high negative charge density due to sulfation. Unfractionated heparin (mean molecular weight between 13000 and 15000) is clinically applied since the 1930s. For many indications, there has been a trend towards use of fractionated low molecular weight heparins (mean molecular weight between 4000 and 7000), manufactured by partial depolymerisation of unfractionated heparin. Low-molecular-weight heparins, such as the most widely used enoxaparin, have lower side effects and a more favourable pharmacokinetics. Safe and effective use of heparins requires maintaining a delicate balance - dosing low enough to minimize the risk of bleeding, yet high enough to treat or prevent thrombosis. Measurement of activated partial thromboplastin time aPTT (the ability to delay clotting) and chromogenic anti-factor Xa assay (inhibition of a specific blood coagulation factor), are the currently accepted practice for laboratory monitoring of unfractionated and low molecular weight heparins, respectively, in patients' blood. Accurate laboratory monitoring has proven to be difficult to achieve. [1-3] Comparability between commcercially available anti-factor Xa assays is poor.

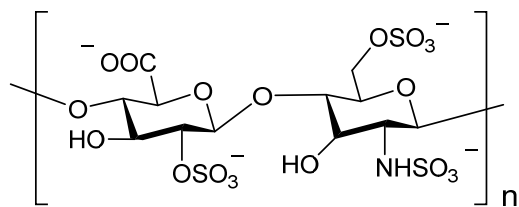

**Scheme 1**. Structure of the major repeating disaccharide unit of heparin.

The drawbacks of the established indirect assays have stimulated the development of direct optical detection methods that rely on the color or fluorescence change of cationic dyes upon binding to polyanionic heparin [4]. Only few such dyes or assay kits, however, are commercially and widely available to reserachers and clinicians.

This study compares the performance of three commercial dyes (assays, respectively) for direct heparin detection in human plasma: The more recently developed fluorescent dye Heparin Red and the traditional metachromatic dyes Azure A and Toluidine Blue.

Heparin Red is a polyamine-functionalized, red-fluorescent perylene diimide (scheme 2). It forms a supramolecular complex with polyanionic polysaccharides in which aggregation of dye molecules results in contact quenching of fluorescence (scheme 3). [5] Strong binding of the polycationic probe to polyanionic heparin appears to be controlled by both electrostatic and aromatic pi-stacking interactions, with formation of charge-neutral aggrgeates. [6] The commercial Heparin Red Kit has been applied to the quantification in the low µg/mL range of unfractionated heparin [7], the low molecular weight heparins enoxaparin [7] and dalteparin [8], heparin octa- and decasacchairides [7], chemically modified heparins [7], heparan sulfate [9], fucoidans [10], carrageenan [8], ulvan [8], dextran sulfate [8] and sulfated hyaluronic acid [8] in human plasma. Heparin Red is used in several drug development projects for pharmacokinetic studies of non-anticoagulant heparins.[11]

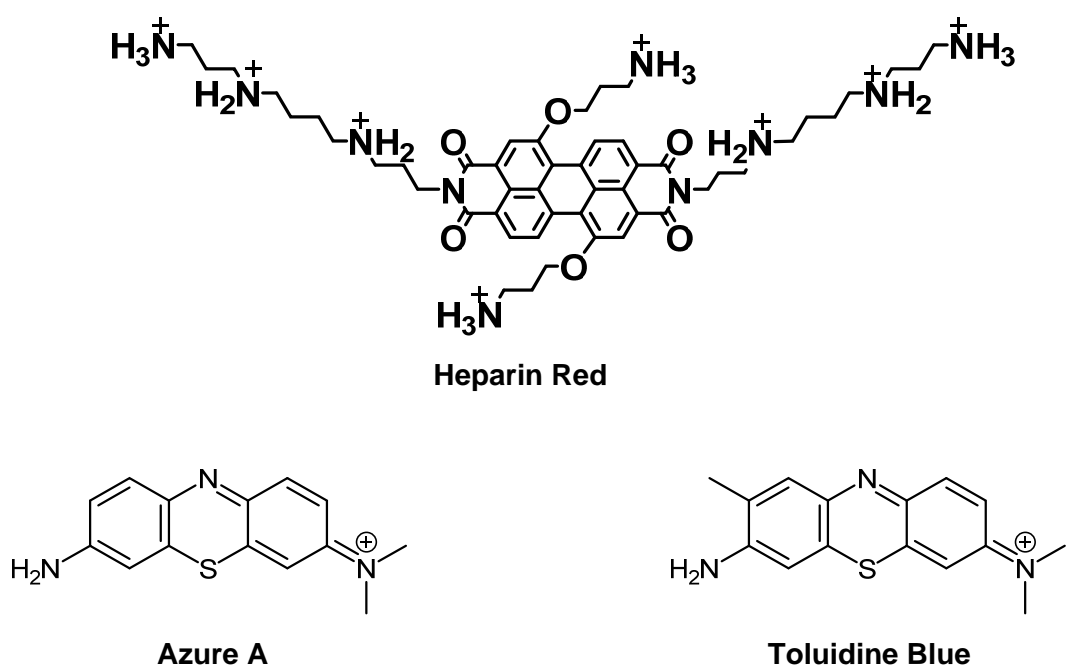

**Scheme 2**. Molecular structures of the polycationic, fluorescent probe Heparin Red, and the monocationic, metachromatic dyes Azure A and Toluidine Blue.

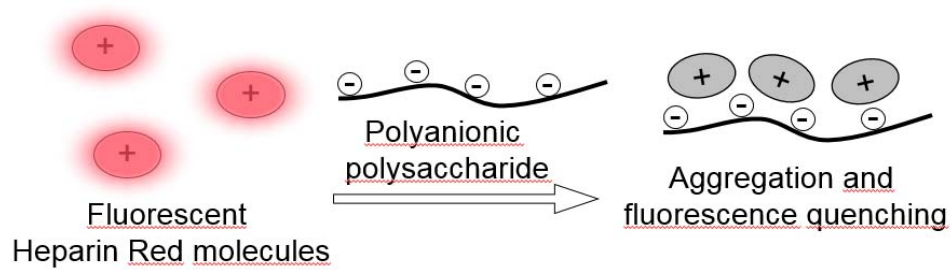

**Scheme 3**. Schematic representation of fluorescence quenching of the molecular probe Heparin Red in the presence of polyanionic polysaccharides.

Azure A and Toluidine Blue (scheme 2) are blue, monocationic dyes of the thiazine family. These dyes show metachromatic behaviour in the presence of sulfated polysaccharides [12,

13], they change color when associating with the target. Applications include staining of sulfated polysaccharides in tissue and (after electrophoresis) in gels, and direct detection in solution if intereference from sample matrix is low. A structurally related thiazine dye, dimethylmethylene blue, is the major component of a commercial kit (Blyscan) that is widely used for sulfated polysaccharide detection. The Blyscan multistep protocol [14] is based on the intentional precipitation of the dye-polysaccharide complex, followed by isolation of the precipitate, re-dissolution of the complex at high salt concentration and colorimetric quantification of the dye component. A similar precipitation- redissolution protocol is followed for sulfated polysaccharide quantification with Alcian Blue [15, 16], a tetracationic copper-phthalocyanine derivative, although this dye does not display metachromatic behaviour. The uses and limitations of these "blue dyes" have been discussed in several reviews and research papers [17-20]. In blood plasma, strong matrix interference complicates the detection of heparin. Only few protocols for the direct determination of unfractionated heparin in plasma or serum samples have been described in the literature.[21-23]

The present study compares the performance and sensitivity of the three commercial dyes Heparin Red, Azure A and Toluidine Blue, using the same heparin-spiked plasma samples and a convenient, mix-and-read microplate protocol.

## Materials and Methods

### Instrumentation

*Absorbance measurements*
Absorbance at 620 or 630 nm was measured with a microplate reader Biotek, EON, (Biotek Instruments, Winooski, VT, USA) using Gen 5 software, version 2.06.10 for data recording.

*Fluorescence measurements*
Fluorescence (Heparin Red® assay) was measured with a microplate reader Biotek Synergy Mx (Biotek Instruments, Winooski, VT, USA), excitation at 570 nm, emission recorded at 605 nm, spectral band width 13.5 nm, read height of 8 mm.

*Microplates*
For absorbance measurements, 96-well microplates, polystyrene, costar Item No 3474, were purchased from Corning Incorporated (NY, USA). For fluorescence measurements (Heparin Red® assay) 96 well microplates, polystyrene, Item No 655076, were purchased from Greiner Bio-One GmbH, Frickenhausen.

**Reagents**

*Dyes and Kits*

The Heparin Red® Kit was a gift from Redprobes UG, Münster, Germany [24]. Kit components: Heparin Red solution, Product No HR001, Lot 01-004, and Enhancer Solution, Product No ES001, Lot 006. Azure A (chloride; dye content 70%) was provided by Sigma-Aldrich GmbH Steinheim (product number 861049, Lot Nr MKBW1530V). Toluidine Blue (tetrachlorozincate) was provided by Sigma-Aldrich GmbH Steinheim (product number 89640, Lot Nr. BCBR0169V.

*Heparins*

Unfractionated heparin sodium salt from porcine intestine mucosa ("heparin"), was sourced as a solid from Sigma-Aldrich GmbH, Steinheim (product number H5515, Lot SLBK0235V, indicated potency 210 IU/mg). We are grateful to Apotheke des Universitätsklinikums Heidelberg for the gift of enoxaparin (Clexane® 20 mg, Sanofi, solution 10000 IU/mL).

*Plasma*

Pooled human plasma was prepared by mixing equal volumes of ten single-donor citrated plasmas of healthy individuals, provided by the Blood Bank of the Institute for Clinical Transfusion Medicine and Cell Therapy Heidelberg (IKTZ). Plasma as well as heparin spiked plasma samples were stored at -20°C.

*Other*

Aqueous solutions were prepared with HPLC grade water purchased from VWR, product No 23595.328. 10% Pluronic® F-68 non-ionic surfactant Item No. 24040032 was purchased from Thermo Fisher Scientific. Dulbecco's Phosphate buffered saline (DPBS) Item No. 14190-094 was purchased from Thermo Fisher Scientific.

**Methods**

*Heparin Red® Kit*

For determination of heparin concentrations in the range <1 IU/mL in in plasma samples, the protocol of the provider for a 96-well microplate assay was followed. A mixture of 100 µL Heparin Red solution and 9 mL Enhancer solution was freshly prepared. 20 µL of the heparin spiked plasma sample was pipetted into a microplate well, followed by 80 µl of the Heparin Red – Enhancer mixture. The microplate was introduced in the fluorescence reader and shaken for 3 minutes, using plate shaking function of the reader (setting "high"), followed by fluorescence measurement within 1 minute. For determination of heparin concentrations up to 10 IU/mL in plasma samples, a mixture of 600 µL Heparin Red solution and 9 mL Enhancer solution was freshly prepared. Otherwise, the protocol was followed as described above.

*Azure A assay, method 1*

For determination of heparin concentrations in plasma samples, the literature protocol [21] for a "4-units assay" was followed, but heparin from porcine mucosa (instead of bovine lung heparin) was used and the protocol adapted for detection in 96-well microplates. A solution of Azure A in water (8 mg / 100 mL, c = 80µg/mL) was freshly prepared. 50 µL of the heparin spiked plasma sample was pipetted into a microplate well, followed by 50 µl of the Azure A solution. The microplate was immediately introduced in the absorbance reader and shaken for 1 minute, using the plate shaking function of the reader (setting 567 cpm), followed by absorbance measurement within 1 minute.

*Azure A assay, method 1a*

For determination of heparin concentrations in plasma samples, the literature protocol [21] for a "10-units assay" was followed, but heparin from porcine mucosa (instead of bovine lung heparin) was used and the protocol adapted for detection in 96-well microplates. A solution of Azure A in water (1 mg / 100 mL, c = 10µg/mL) was freshly prepared. 10 µL of the heparin spiked plasma sample was pipetted into a microplate well, followed by 200 µl of the Azure A solution. The microplate was immediately introduced in the absorbance reader and shaken for 1 minute, using the plate shaking function of the reader (setting 567 cpm), followed by absorbance measurement within 1 minute.

*Azure A assay, method 2*

For determination of heparin concentrations in plasma samples, a protocol described in a patent [22] was followed, but the commercial Azure A sample was not purified, human plasma was used (instaed of horse serum) and the protocol adapted for detection in 96-well microplates. A solution of Azure A (14.6 µg / mL) in 1% aqueous Pluronic (commercial product was diluted 10-fold with water) was freshly prepared. 10 µL of the heparin spiked plasma sample was pipetted into a microplate well, followed by 100 µl of the Azure A solution. The microplate was immediately introduced in the absorbance reader and shaken for 1 minute, using the plate shaking function of the reader (setting 567 cpm), followed by absorbance measurement within 1 minute.

*Toluidine Blue assay*

For determination of heparin concentrations in plasma samples, a literature protocol [23] was followed, but sodium heparin (instead of lithium heparin) and citrated human plasma (instead of EDTA human plasma) were used and the protocol adapted for detection in 96-well microplates. A solution of Toluidine Blue (100 µg / mL) in phosphate buffered saline (DPBS) was freshly prepared. 50 µL of the heparin spiked plasma sample was pipetted into a microplate well, followed by 50 µl of the Toluidine Blue solution. The microplate was immediately introduced in the absorbance reader and shaken for 1 minute, using the plate shaking function of the reader (setting 567 cpm), followed by absorbance measurement within 1 minute.

*Preparation of spiked plasma samples*

Plasma samples containing defined concentrations of heparins were prepared as follows: Aqueous solutions (10 vol%) of unfractionated heparin and enoxaparin, respectively, were added too pooled human plasma to achieve a concentration of 10 IU/mL. Heparin concentrations required for the detections were adjusted by further dilution of this 10 IU/mL stock solution with the same plasma. The spiked plasma samples were stored at -20°C and thawed at room temperature before use.

*Data analysis*

Data were analyzed using Excel (Microsoft Office 10). Linear regression "through origin" ( with y-intercept set to 1) was applied to the response curves in figure 2. Coefficients of determination ($R^2$) in table 1 were determined by linear regression.

## Results and discussion

While the Heparin Red, Azure A and Toluidine Blue assay actually determine a mass concentration of heparin (the "chemical" heparin) rather than a biological activity, we refere in this work to IU (international units) of heparin for better comparability with published data for the metachromatic dyes and clinically relevant heparin levels. The heparin batch used in this work had a potency of 210 IU/mg solid, i.e. 1 IU corresponds to 4,8 µg.

The Heparin Red Kit was applied either as suggested by the protocol of the supplier, or adapted to higher heparin levels by using a higher concentrations of the dye (see "Materials and Methods" for details). For Azure A, we compared three different published protocols (method 1, 1a and 2) for heparin detection. In method 1 and 1a [21], an aqueous solution of the dye and the heparin containing plasma sample are mixed in either a 1:1 ratio (for heparin levels up to 4 IU/mL, method 1) or 1:9 ratio (for heparin levels up to 10 IU/mL, method 1a) in a cuvette and dye absorbance at 620 nm is measured with a photometer. Azure A, Method 2, was described in a patent [22]: an aqueous solution of the dye containing an anionic surfactant is mixed with a heparin containing sample of horse serum at a 10:1 ratio in a cuvette and the absorbance spectrum recorded with a photometer. A protocol using Toluidine Blue [23] suggests mixing of equal amounts of a dye solution in phosphate buffered saline and heparin containing plasma sample, followed by absorbance readout with a Nanodrop photometer at 630 nm.

We adapted these protocols for microplates, i.e. dye solution and plasma sample were mixed in 96-well microplate wells and absorbance read with a microplate reader. For Azure A, method 2, the serum was replaced by a human plasma sample. The linear relation between dye concentration and absorbance readout of the plate reader in the relevant concentration range of Azure A is shown in figure 1. Solutions of Toluidine Blue in the range 0-40 µg/mL also displayed good linearity of absorbance (data not shown).

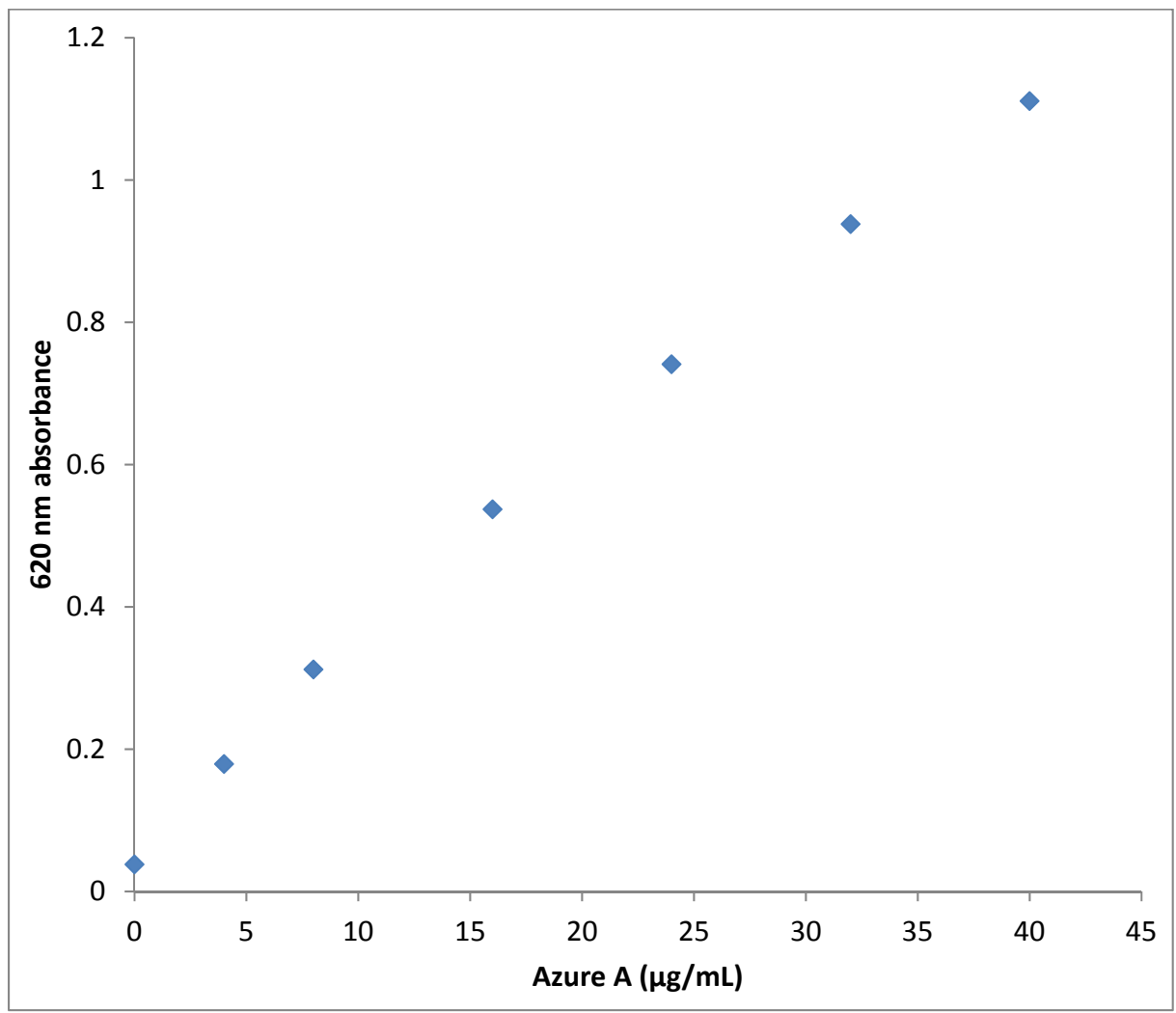

**Figure 1**. Linear relation between Azure A mass concentration (range 0-40 µg/mL) and 620 nm absorbance of aqueous solutions of Azure A. Absorbance (arbitrary units) measured in microplate wells (sample volume 100 µL) with a microplate reader.

**Quantification of unfractionated heparin and enoxaparin at concentrations <1 IU/mL in human plasma**

In the vast majority of clinical heparin applications, the recommended plasma level is < 1 IU/mL. More precisely, the range for therapeutic dosing is 0.3 – 0.7 IU/mL for unfractionated heparin and 0.5 – 1.0 IU/mL for enoxaparin, and for prophylactic doses 0.1 – 0.4 IU/mL for heparin and 0.2 – 0.4 IU/mL for enoxaparin. Therefore, the first series of heparin determinations (figure 2) focuses on concentrations < 1 IU/mL plasma. The fact that all dyes respond to heparin by a decrease of optical signal (fluorescence at 610 nm or absorbance at about 620 nm, respectively) facilitates the comparison of their performance in heparin detection. The normalized response of Heparin Red, Azure A (method 1 and 2) and Toluidine Blue to the same heparin spiked human plasma samples is shown in figure 2.

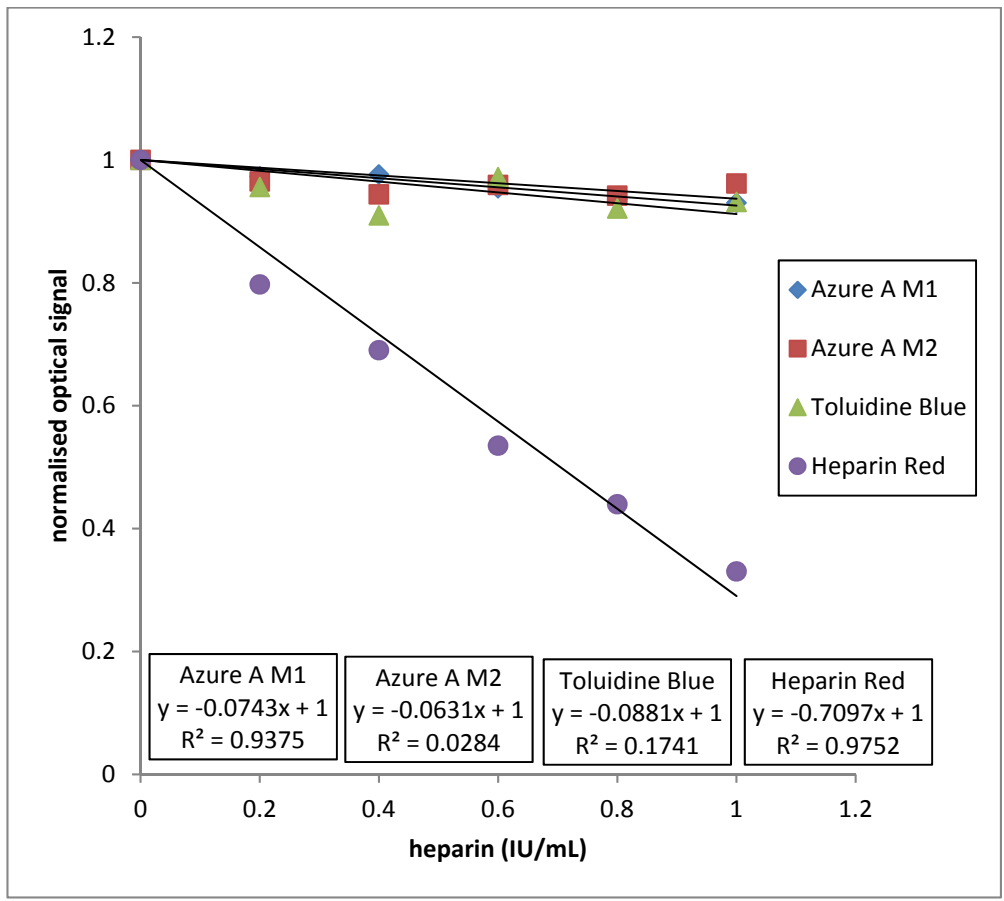

**Figure 2**. Normalized heparin dose response curves of the four assays Heparin Red, Azure A (M1 = method 1, M2 = method 2), and Toluidine Blue. "Optical signal" refers in case of Heparin Red to 610 nm fluorescence, for Azure A assays to 620 nm absorbance and for Toluidine Blue assay to 630 nm absorbance. Heparin in IU/mL refers to the concentration in spiked pooled normal plasma samples. Manually performed microplate assays, as described in the "Materials and Methods" section. Averages of duplicate determinations.

While the fluorescence signal of Heparin Red displays a strong decrease in the range 0-1 IU/mL, the other dyes show only poor response. Heparin levels <1 IU/mL have actually not been measured by the literature protocols for Azure A, method 1, and Toluidine Blue. The lowest measured plasma concentrations were 1 IU/mL for Azure A, method 1 [21] and 1.7 IU/mL for Toluidine Blue [23]. Azure A, method 2, described in a patent [22], claims a detection limit of 0.1 IU/mL heparin in horse serum, derived from changes in the area under the absorption spectra. In our hands, however, this protocol has the poorest detection limit and linearity (table 1) of all methods when applied to heparin-containing plasma samples in a microplate, independent of how the optical signal change is monitored.

The detection and quantification limits of the assays (table 1) were determined based on signal-to-noise [25], by relating extrapolated response (linear regression "through origin", y intercept set to 1) to the standard deviation of blank samples ($\sigma_{blank}$) without heparin. The limit of detection (LOD) was calculated as LOD = 3 $\sigma_{blank}$ / S (S= slope of response curve, see figure 1) and the limit of quantification as LOQ = 10 $\sigma_{blank}$ / S.

| Assay | Heparin Red | Azure A M1 | Azure A M2 | Toluidine Blue |
|---|---|---|---|---|
| $\sigma_{blank}$ (n=8) | 0.014 | 0.013 | 0.020 | 0.025 |
| $r^2$ | 0.98 | 0.94 | 0.03 | 0.17 |
| LOD (IU/mL) | 0.06 | 0.54 | 0.94 | 0.85 |
| LOQ (IU/mL) | 0.20 | 1.8 | 3.1 | 2.8 |

**Table 1**. $\sigma_{blank}$, coefficient of determination ($r^2$), limit of detection (LOD) and limit of quantification (LOQ) of heparin in human plasma by the four assays Heparin Red, Azur A M1, Azure A M2 and Toluidine Blue. $\sigma_{blank}$ is the standard deviation of the normalized optical signal of a heparin-free dye-plasma mixture. $r^2$ is the coefficient of determination obtained from linear regression (figure 2). LOD = 3 $\sigma_{blank}$ / S (S= slope of response curve, see figure 2). LOQ = 10 $\sigma_{blank}$ / S.

When compared on the basis of optical signal change (figure 2), the Heparin Red assay is at least 8 times more sensitive to heparin than the metachromatic dyes. When compared on the basis of detection and quantification limits (table 1), Heparin Red is at least 9 times more sensitive than the metachromatic dyes. Quantification limits of the latter lie outside the 0-1 IU/mL concentration range. In addition, the Heparin Red assay has the best linear correlation (table 1) between heparin concentration and optical signal.

Response of the assays to the low molewcular weight heparin enoxaparin (figure 3) follows similar trends as observed for unfractionated heparin. The decrease of optical signals is more pronounced since in the heparin and enoxaparin solutions of the same potency (expressed in IU/mL), the mass concentration of enoxaparin is significantly higher, and the dye assays respond to mass concentration rather than potency (compare reference [7]).

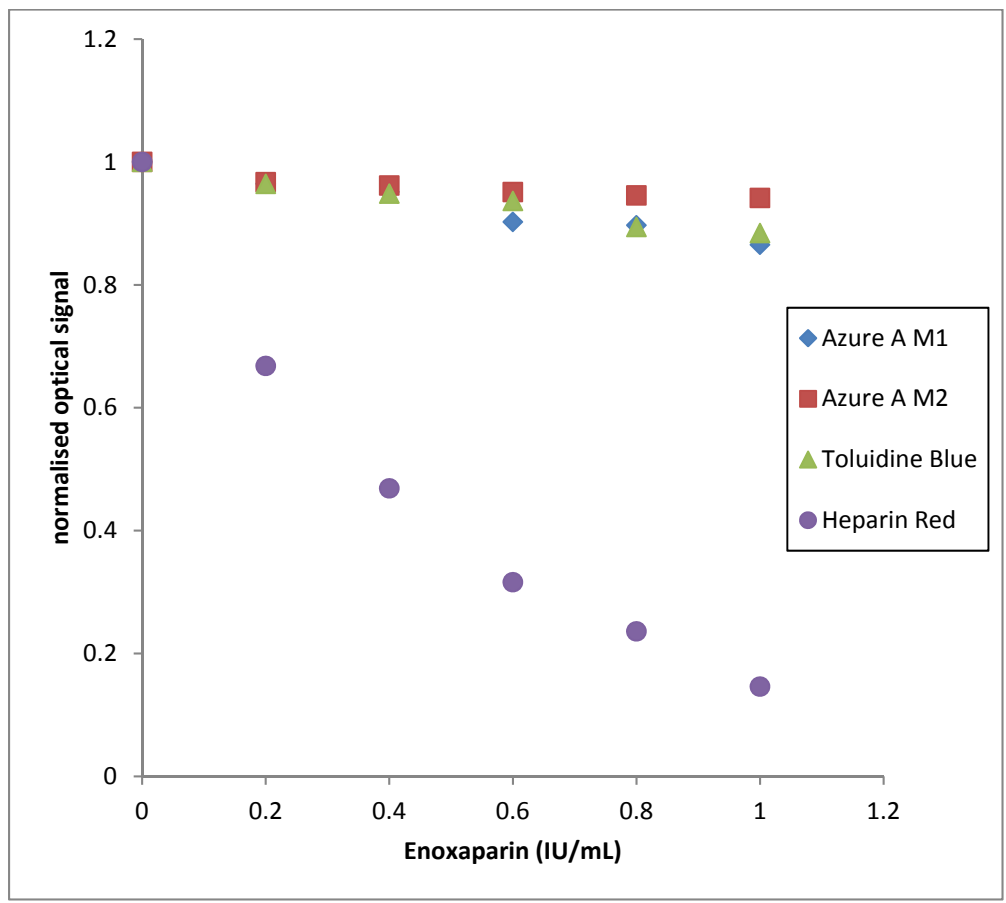

**Figure 3.** Normalized enoxaparin dose response curves of the four assays Heparin Red, Azure A (M1 = method 1, M2 = method 2), and Toluidine Blue. "Optical signal" refers in case of Heparin Red to 610 nm fluorescence, for Azure A assays to 620 nm absorbance and for Toluidine Blue assay to 630 nm absorbance. Enoxaparin in IU/mL refers to the concentration in spiked pooled normal plasma samples. Manually performed microplate assay, as described in the "Materials and Methods" section. Averages of duplicate determinations.

**Quantification of unfractionated heparin at concentration up to 10 IU/mL in human plasma**

In specific clinical situations such as cardiopulmonary bypass surgery, application of unfractionated heparin concentrations at blood levels up to 10 IU mL is adequate. The Heparin Red assay is readily adapted to this concentration range (figure 4) by increasing the dye concentration, see "Materials and Methods" for details. The metachromatic dye assays are also suitable for heparin quantification in this extended concentration range, although their response is more than two-fold weaker compared with Heparin Red. A good match with literature data is observed for Azure A, method 1 (2 and 4 IU/mL heparin), while our protocol provides a weaker response for Azure A, method 1a and Azure A, method 2, and a stronger response for the Toluidine Blue assay. This could be related to different heparins and matrices used in the literature reported assays (see "Materials and Methods" for details).

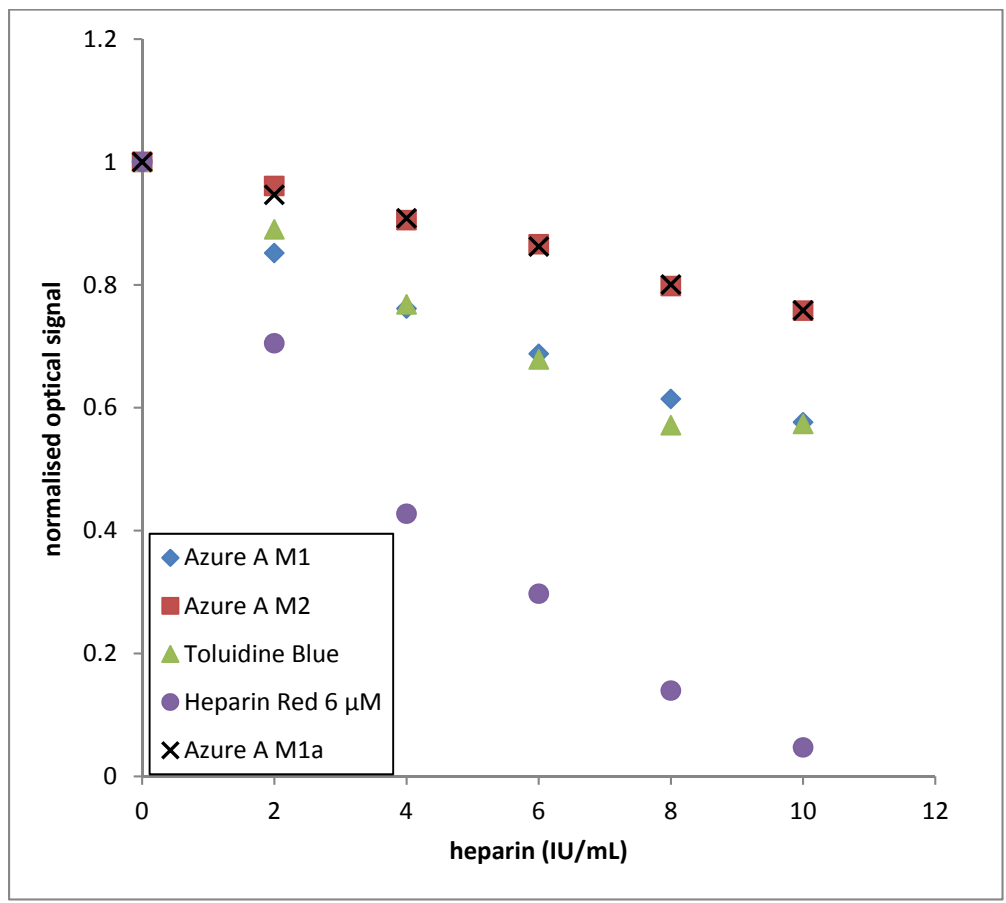

**Figure 4.** Normalized heparin dose response curves of the five assays Heparin Red, Azure A (M1 = method 1, M1a = method 1a, M2 = method 2), and Toluidine Blue. "Optical signal" refers in case of Heparin Red to 610 nm fluorescence, for Azure A assays to 620 nm absorbance and for Toluidine Blue assay to 630 nm absorbance. Heparin in IU/mL refers to the concentration in spiked pooled normal plasma samples. Manually performed microplate assay, as described in the "Materials and Methods" section. Averages of duplicate determinations.

## Conclusion

Heparins are widely used anticoagulant drugs. In clinical practice, blood levels of heparins are monitored indirectly by their effect on coagulation factors, although accurate laboratory monitoring has proven to be difficult to achieve. Direct detection of heparins by cationic dyes that change absorbance or fluorescence upon binding of polyanionic heparin offers an alternative tool for quantification. Only few such dyes or assay kits, however, are commercially and widely available to researchers and clinicians. The present study compares the performance and sensitivity of three commercial dyes for the direct quantification of heparins in human plasma: two traditional metachromatic dyes, Azure A and Toluidine Blue, and the more recently developed fluorescent dye Heparin Red. In the clinically most relevant concentration range below 1 IU per mL, only Heparin Red is a useful tool for determination of heparins. It is at least 9 times more sensitive than the metachromatic dyes which can not be reliably quantify the heparins in this concentration range if the limit of quantification is defined as 10 $\sigma_{blank}$ / S ($\sigma_{blank}$: standard deviation of blank; S: slope of response curve). Higher heparin levels of 2 -10 IU/mL can be determined by all dyes, Heparin Red being the most sensitive.

**Conflict of interest.** R. Krämer holds shares in Redprobes UG, Münster, Germany. Other authors: No conflict of interest.